# Simultaneous off-axis multiplexed holography and regular fluorescence microscopy of biological cells

YOAV N. NYGATE, GYANENDRA SINGH, ITAY BARNEA, AND NATAN T. SHAKED*

*Tel Aviv University, Faculty of Engineering, Department of Biomedical Engineering, Tel Aviv 69978, Israel.*
*Corresponding author: nshaked@tau.ac.il*

**We present a new technique for obtaining simultaneous multimodal quantitative phase and fluorescence microscopy of biological cells, providing both quantitative phase imaging and molecular specificity using a single camera. Our system is based on an interferometric multiplexing module, externally positioned at the exit of an optical microscope. In contrast to previous approaches, the presented technique allows conventional fluorescence imaging, rather than interferometric off-axis fluorescence imaging. We demonstrate the presented technique for imaging fluorescent beads and live biological cells.**

Throughout the years, various microscopic methods have been developed in order to enhance contrast and differentiate between relevant and irrelevant information when imaging biological samples. Specifically, live biological cells are almost entirely transparent when imaged *in vitro* using bright-field microscopy, making it difficult to discern cell contents. The most commonly used method for enhancing biological sample imaging contrast is staining or labeling the sample. In fluorescence microscopy, cell labeling is implemented by introducing fluorophores into the sample, which then proceed to bond to the targeted membranes or molecules within the sample. This technique is not restricted to the imaging of fixed cells and can be applied in live-cell imaging as well.

An alternative approach used to enhance the contrast of biological samples is phase microscopy. Phase microscopy takes advantage of the fact that the amount of delay accumulated by the illumination light as it passes through the different regions in the sample is dependent on the refractive indices and sample thicknesses at each point in the sample. In contrast to Zernike's phase contrast microscopy and differential interference contrast (DIC) microscopy, quantitative phase microscopy (QPM) enables the acquisition of contrast at all spatial points in the sample image. Furthermore, the contrast obtained by QPM is quantitative; the optical path delay (OPD) of the light as it propagates through each spatial point in the sample is obtained, producing a quantitative OPD value for each spatial point in the sample image, as opposed to the non-quantitative values acquired using Zernike's phase contrast microscopy or DIC, as well as label-based approaches, such as fluorescence microscopy. The OPD value obtained at each point is equal to the product of the sample thickness at that point in the sample and the integral refractive index along this thickness. However QPM, contrary to label-based microscopy, does not enable molecular specificity, meaning that it is not possible to selectively image specific cell organelles. This is due to the fact that the QPM contrast mechanism is based on the cell refractive index, which may not be unique to the organelles of interest. Simultaneously obtaining both molecular specificity using fluorescence microscopy and quantitative contrast using QPM enables unique cell characterizations, such as separate quantitative phase investigation of certain organelles. Several previous works have proposed techniques for measuring fluorescent emission together with quantitative phase imaging in order to gain molecular specificity in biological cell imaging [1-5]. However, these approaches use two different imaging channels and two different cameras, one for QPM and the other for fluorescence microscopy. This leads to complex microscopy systems and difficulties in registering the two images from the two cameras when processing the data.

QPM of dynamic biological cells is typically implemented using off-axis holography, which captures the complex wavefront of the sample with a single camera exposure. This is done by inducing a small angle between the sample and the reference beams, creating the off-axis interference pattern of the hologram. Phase reconstruction from this single off-axis interference pattern is possible as, in the spatial frequency domain, there is full separation between the auto-correlation term that originates from the sample and reference beam intensities and each of the cross-correlation terms, each of which contains the complex wavefront of the sample. This spatial frequency separation typically occurs along a single axis, which allows the encoding of more information along the other axes as well. This can be done by optically multiplexing several holograms with different interference fringe orientations into a single hologram and fully reconstructing each of the complex wavefronts encoded. Each of these holograms can contain additional data on the imaged sample, meaning that multiplexing allows the recording of more information with the same camera pixels. This is beneficial for highly dynamic samples as more data can be recorded in a single exposure, as opposed to acquiring the data sequentially in multiple camera exposures between which the dynamic sample may change significantly.

Using this off-axis holographic multiplexing principle, we have previously presented the technique of off-axis interferometry with doubled imaging area (IDIA) [6]. In this technique, we optically compress two off-axis holographic images into a single camera image and gain an extended field of view. This is done by an external holographic module capable of projecting one reference beam and two sample beams onto the camera simultaneously. Each sample beam contains a different field of view of the sample, and corresponds to a different off-axis interference fringe orientation, preventing overlap of the cross-correlation terms in the spatial frequency domain. This approach was later extended to imaging two holograms of different wavelengths in order to perform two-wavelength phase unwrapping [7]. Chowdhury et al. [8] used the principle of off-axis holographic multiplexing in order to obtain QPM and fluorescence microscopy images simultaneously on the same camera, without image registration issues. They implemented a diffraction-based interferometric phase microscopy system that enabled them to perform white-light holography, allowing them to create off-axis holograms of fluorescence images. However, this setup was limited to white-light interferometers due to the restrictions imposed by the very short coherence length of the fluorescent emission and the requirement of obtaining off-axis fluorescence holograms.

In the present paper, we propose an external off-axis interferometric module, termed as the DC-free off-axis τ (DCF-τ) interferometer, which can be attached to the output of a conventional fluorescence microscope. This module enables a single exposure QPM acquisition with off-axis holography, even if illuminated with coherent or low-coherence light sources, together with conventional (in-line) fluorescence imaging provided by a commercial fluorescence microscope. Holographic multiplexing on the same digital camera is used to simultaneously acquire the sample intensity in an additional off-axis holographic channel in order to subtract it from the overlapping fluorescence image. It should be noted that due to the incoherent nature of the fluorescent signal, the fluorescence image that is projected onto the output camera does not produce a hologram, thus no fluorescence interference pattern is generated.

Figure 1(a) shows a simplified scheme of the imaging system used, with the proposed DCF-τ module connected at its output. The imaging system is a commercial inverted microscope (IX83, Olympus), which contains an epi-illuminance fluorescence imaging channel. For low-coherence off-axis holography, we illuminated the microscope input with a supercontinuum laser source (SuperK Extreme, NKT), coupled to an acousto-optic tunable filter, AOTF (SuperK SELECT, NKT), which emits a wavelength bandwidth of 633 ± 2.5nm. This light is reflected by mirror M1 and enters the inverted microscope. Inside the microscope, the beam passes through the sample S, and is magnified by microscope objective MO (Olympus UPLFLN, 40×, 0.75 NA). Simultaneously, white light is emitted by a mercury lamp (U-HGLGPS 130W Mercury burner, emission wavelength 360-770 nm), a component contained in the microscope, and is filtered by excitation filter ExF (350 nm ± 25 nm) and reflected by dichroic mirror DM (longpass with cutoff at

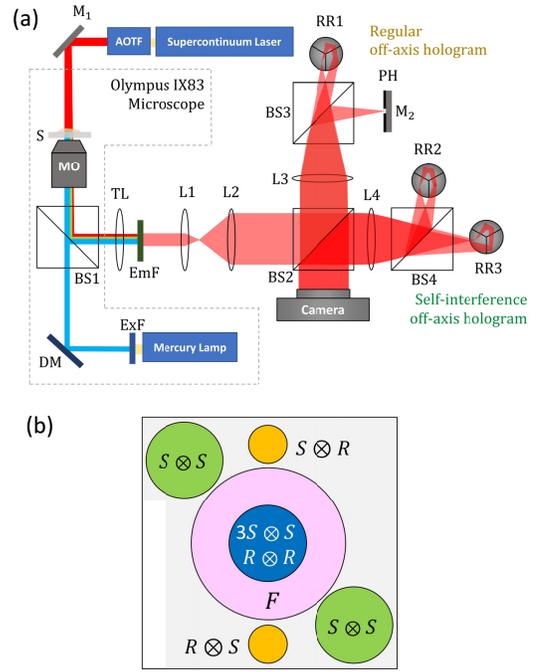

Fig. 1. (a) A scheme of the optical system, containing an inverted microscope, with the DCF-τ module positioned at its output. AOTF, acousto-optic tunable filter; BS1 – BS4, beam-splitters; DM, dichroic mirror; EmF, emission filter; ExF, excitation filter; L1 – L4, lenses; M1, M2, mirrors; MO, microscope objective; PH, pinhole; RR1 – RR3, 3-mirror retroreflectors; S, sample; TL, tube lens. (b) Scheme of the Fourier transform of the off-axis multiplexed hologram recorded, outlining the fully separable terms.

400 nm). The resulting excitation light then passes through beam-splitter BS1 and MO and illuminates the sample. The labeled sample then proceeds to fluoresce and emits green light that passes through BS1 together with the reflection of the blue fluorescence excitation light and the red QPM light from the supercontinuum laser. The three beams are then reflected toward the tube lens TL (f = 200 mm), which projects the beams onto the output image plane of the microscope where longpass emission filter EmF (cutoff at 496 nm) is positioned. This allows only the green fluorescence emission light and the red coherent light to enter the external multiplexing module.

In the external DCF-τ module, the beams are first magnified by a 4f system composed of lenses L1 (f = 45 mm) and L2 (f = 150 mm) and their intensities are then split at beam-splitter BS2 and directed into two different interferometers, one (shown at the top of Fig. 1(a)) that generates an off-axis hologram between the sample beam and reference beam, and the second (shown on the right of Fig. 1(a)) that generates an off-axis hologram of the sample intensity. The first interferometer is a modification of our external off-axis τ interferometer module [9]. In this τ module the beam is first Fourier transformed by lens L3 (f = 150 mm) and then split into two separate beams by beam-splitter BS3. The first of these beams is spatially filtered by the pinhole PH and is reflected back by mirror M2, thus creating a reference beam. The second beam created by BS3 is reflected by 3-mirror retroreflector RR1 (SSI Optics, 25.4mm Clear Aperture Aluminum Retroreflector, 3 Arc sec), creating an off-axis angle between the two beams when they are projected onto the digital camera (Basler, acA2440-75um), after passing once again through lens L3 and beam-splitter BS2. The second interferometer contains lens L4 (f = 100 mm) that Fourier transforms the second beam produced by BS2. It also contains beam splitter BS4 and two 3-mirror retroreflectors, RR2 and RR3, creating two sample beams with an off-axis angle between them. These beams are recombined after they pass once again through lens L4 and are reflected onto the camera by beam-splitter BS2. Since we use 3-mirror retroreflectors, retaining the solid angles of the reflected beams, the two sample beams arrive at the camera with an accurate spatial overlap between them [10]. In addition, even though the fluorescent signal is split into three copies of itself, in the end it is recombined by the beam-splitters with an exact spatial overlap, resulting in a single fluorescent signal with no shift between the recombined copies. Overall, on the camera we obtain a multiplexed off-axis hologram, containing two holograms with different fringe orientations – a conventional off-axis hologram of the sample and an off-axis self-interference hologram. In addition, the regular fluorescence image is projected onto the camera. Figure 1(b) shows a theoretical scheme of the spatial frequencies obtained. Since the self-interference off-axis hologram retains a copy of the auto-correlation term, located diagonally off-axis, we can allow in-line overlap between the auto-correlation term of the regular off-axis interference hologram and that of the fluorescence image, both located at the origin of the spatial frequency domain. The in-line fluorescence image can be extracted by subtracting the auto-correlation term obtained off-axis in the spatial frequency domain. The phase image can then be obtained as usual, from one of the cross-correlation terms [11].

The conventional hologram created on the camera can be mathematically formulated as follows:

$$I_1 = |S' + R|^2 = |r(x,y)|^2 + |s(x,y)|^2$$
$$+ r^*(x,y)s(x,y)\exp\left(-jk(\varphi_s - \varphi_r + y\sin\theta_y)\right)$$
$$+ r(x,y)s^*(x,y)\exp\left(jk(\varphi_s - \varphi_r - y\sin\theta_y)\right). \quad (1)$$

where $S'(x,y) = s(x,y)\exp(-jk\varphi_s)\exp(-jky\sin\theta_y)$ is the sample beam slightly inclined at an off-axis angle of $\theta_y$ along the $y$ axis, $R(x,y) = r(x,y)\exp(-jk\varphi_r)$ is the reference beam, and $\varphi_s$ and $\varphi_r$ are the sample and reference beam phases, respectively. In the spatial frequency domain, the last two terms of Eq. 1 are the cross-correlation terms, each containing the complex wavefront of the sample. The first two terms in this equation are the auto-correlation terms, centered at the origin of the spatial frequency domain. These terms will overlap with the Fourier transform of the fluorescence intensity $F$ and therefore $F$ normally cannot be reconstructed. We therefore simultaneously acquire another off-axis hologram of the interference between two copies of the sample beam, mathematically formulated as follows:

$$I_2 = |S + S''|^2 = 2|s(x,y)|^2 + |s(x,y)|^2\exp(-jkv)$$
$$+ |s(x,y)|^2\exp(jkv). \quad (2)$$

where $S(x,y) = s(x,y)\exp(-jk\varphi_s)$ is the sample beam, $S''(x,y) = s(x,y)\exp(-jk\varphi_s)\exp(-jkv)$ is the sample beam slightly inclined at an off-axis angle along the diagonal line $y = -x$, and $v = x\sin\theta_x + y\sin\theta_y$. The overall multiplexed hologram recorded by the camera in a single exposure is mathematically formulated as: $I_M = I_1 + I_2 + F$. By cropping one of the off-axis shifted auto-correlation terms, and subtracting it three times from the central term, we can obtain the fluorescence image with a bias of the reference beam intensity. Note that if the reference beam is constant, its spatial frequencies are located in a single point at the center of the spatial frequency domain, thus this bias is negligible. In addition to the fluorescence image, it is still possible to reconstruct the quantitative phase information of the sample in the conventional way by cropping one of the y-shifted cross-correlation terms.

We experimentally implemented the system shown in Fig. 1(a). First, we imaged fluorescent beads (6 µm melamine resin microbeads, refractive index of 1.68, GFP labeled, Fluka Analytical) in oil immersion medium (Zeiss), with refractive index of 1.52. On the camera, we obtained three signals simultaneously: the regular off-axis hologram of the sample (sample plus reference), the self-interference off-axis hologram of the sample (sample plus sample), and the regular, in-line fluorescence image. At first, each of these three terms was tested separately. By blocking the self-interference hologram arm and the fluorescence, the conventional hologram was obtained, as shown in Fig. 2(a). Then, by blocking the regular off-axis hologram arm and the fluorescence, the self-interference hologram was obtained, as shown in Fig. 2(b).

Note that the off-axis fringes in this self-interference hologram do not curve as they pass through the bead, in contrast to the hologram shown in Fig. 2(a) that retains the phase information. Finally, by completely blocking the laser illumination (both hologram arms), a regular fluorescence image was obtained, as shown in Fig. 2(c). Note that to allow simultaneous acquisition, we equalized the intensities of the coherent source with that of the fluorescent signal. Also note that there is intensity loss in the beam splitters combining the signals.

Finally, by illuminating with both the laser and the white-light source, without blocking any arms in the system, a multiplexed off-axis image hologram was recorded, as shown in Fig. 2(d). This multiplexed hologram was then digitally processed in order to reconstruct both the OPD map and the fluorescence image. First, the multiplexed hologram was digitally Fourier transformed, resulting in a spatial frequency domain containing two off-axis cross-correlation terms, two off-axis auto-correlation terms, and a central auto-correlation term that overlaps with the fluorescence auto-correlation term, as shown in Fig. 2(e). It can be seen that this spatial frequency domain matches the theoretical domain shown in Fig. 1(b). Next, one of the off-axis auto-correlation terms representing the self-interference and one of the cross-correlation terms representing the complex wavefront of the sample were extracted. The central terms were then isolated and inverse Fourier transformed back to the image domain, resulting in an image that contains the sample intensity of both the coherent channel and the fluorescence image, with the result shown in Fig. 2(f). In addition, the off-axis auto-correlation term was inverse Fourier transformed as well, resulting in the image shown in Fig. 2(g). Finally, in order to decouple the fluorescent signal from the auto-correlation terms, the off-axis auto-correlation image was subtracted three times from the image produced from the central terms, as described previously. This resulted in the reconstruction of the fluorescence image of the sample shown in Fig. 2(h) [11], with a clear resemblance to the original fluorescence image of the bead shown in Fig. 2(c). In order to reconstruct the OPD map of the sample using conventional off-axis holographic processing, one of the cross-correlation terms was cropped and then inverse Fourier transformed.

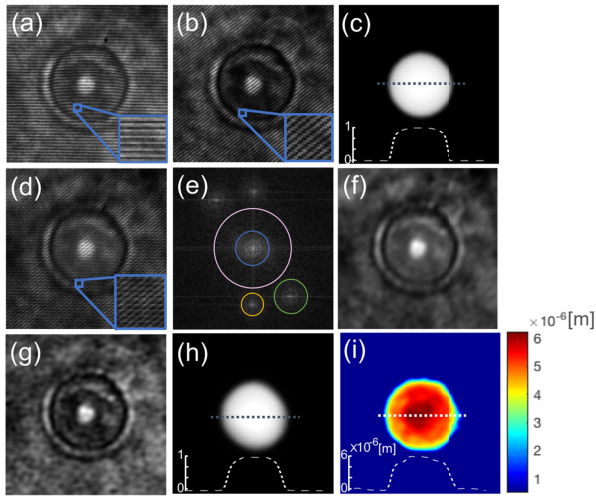

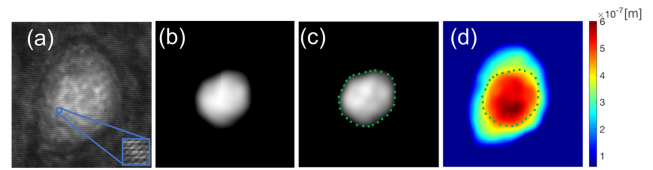

Fig. 2. (a-c) Fluorescent bead separate-channel holograms and fluorescence image. (a) Regular off-axis hologram (sample plus reference), with an enlargement displaying the fringe orientation. (b) Self-interference off-axis hologram (sample plus sample), with an enlargement displaying the fringe orientation. (c) The original fluorescence image (obtained by blocking the illumination of the laser). (d-i) Simultaneous multiplexed hologram/fluorescence imaging. (d) The multiplexed hologram, with an enlargement showing the orientations of the fringes (a combination of horizontal and diagonal fringes). (e) Absolute value of the spatial frequency domain of the multiplexed hologram. Blue circle indicates the auto-correlation terms, pink circle indicates the in-line fluorescence term, green circle indicates one of the off-axis auto-correlation terms obtained from the self-interference hologram, and orange circle indicates one of the cross-correlation terms obtained from the regular hologram. (f) The inverse Fourier transform of the cropped central terms, including the intensities of the coherent laser channels and the overlapping fluorescence image. (g) The inverse Fourier transform of the cropped off-axis auto-correlation term originating from the self-interference hologram. (h) The reconstructed fluorescence image. (i) The reconstructed height map. Colorbar represents height values in m.

Next, we applied a digital 2D phase unwrapping algorithm on the phase argument of the resulting complex wavefront, as described in Ref. [11] and edge steepness compensation. Lastly, dividing the OPD map by $\Delta n$, resulted in the height map shown in Fig. 2(i).

Finally, we used the proposed system for imaging SW480 colon cancer cells labeled with acridine orange nucleus stain, resulting in the off-axis multiplexed hologram shown in Fig. 3(a). Then, the laser signal was blocked and the pure fluorescence image shown in Fig. 3(b) was captured, for validation purposes. The multiplexed hologram was then digitally processed as described earlier, resulting in the reconstructed fluorescence image shown in Fig. 3(c), with a clear resemblance to the original fluorescence image shown in Fig. 3(b), and the reconstructed OPD map shown in Fig. 3(d). The fluorescence image shown in Fig. 3(c) was then used to locate the area of the nucleus on the OPD map shown in Fig. 3(c), as indicated by the broken green line.

To test the system in a highly dynamic situation, we acquired a video of an off-axis multiplexed hologram of an unfixed SW480 colon cancer cell, which can be seen vibrating due to environmental conditions and internal biological processes. The results can be seen in Visualization 1.

Fig. 3. (a) The multiplexed hologram of an SW480 cancer cell, with an enlargement showing the fringe orientations. (b) The original fluorescence image of the cell. (c) The final reconstructed fluorescence image of the cell's nucleus. (d) The reconstructed OPD map of the cell, with marking of the nucleus area detected from the fluorescence image. Colorbar represents OPD values in meters. See Visualization 1 for dynamic reconstruction.

To conclude, the DCF-τ module presented here leaves empty space around the center of the spatial frequency domain due to the parallel self-interference channel that is subtracted from the central terms there. This allows us to use this empty space to record a regular fluorescence image, without image registration problems that might occur when using two different cameras. Note, however, that since all signals share the same grayscale dynamic range on the camera, samples with high absorption cannot be imaged with the proposed system. Additionally, as the hologram and the fluorescence image must be acquired together, this technique is suitable for highly fluorescent samples, such as the nucleus staining demonstrated here, and is not suitable for measuring weak fluorescent signals. Also note that there is still empty space in the spatial frequency domain that may be used for inserting more parallel imaging channels.

We believe that the proposed technique can aid in cytometry studies that require both quantitative imaging and molecular specificity for highly dynamic samples. In general, the simplicity of the setup, its portability and the lack of image registration problems in the integrated data processing are expected to make this approach attractive for biological and medical applications.

**Funding.** Horizon 2020 European Research Council (678316).

# References with titles